# Fluorination Increases Hydrophobicity at the Macroscopic Level but not at the Microscopic Level

Weishuai Di (邸维帅)[1], Xin Wang (王鑫)[1], Yanyan Zhou (周艳艳)[1], Yuehai Mei (梅岳海)[2], Wei Wang (王炜)[2, 3, 4, *], and Yi Cao (曹毅)[1, 2, 3, 4, †]

[1]*Oujiang Laboratory (Zhejiang Lab for Regenerative Medicine, Vision and Brain Health), Wenzhou Institute, University of Chinese Academy of Sciences, Wenzhou 325001, China*

[2]*Collaborative Innovation Center of Advanced Microstructures, National Laboratory of Solid State Microstructure, Department of Physics, Nanjing University, Nanjing 210093, China*

[3]*Institute for Brain Sciences, Nanjing University, Nanjing, 210093, China and*

[4]*Chemistry and Biomedicine Innovation Center, Nanjing University, Nanjing 210093, China*

∗ Corresponding author: wangwei@nju.edu.cn

† Corresponding author: caoyi@nju.edu.cn

Hydrophobic interactions have been studied in detail in the past based on hydrophobic polymers, such as polystyrene (PS). Because fluorinated materials have relatively low surface energy, they often show both oleophobicity and hydrophobicity at the macroscopic level. However, it remains unknown how fluorination of hydrophobic polymer influences hydrophobicity at the microscopic level. In this work, we synthesized PS and fluorine-substituted PS (FPS) by reversible addition- fragmentation chain transfer polymerization method. Contact angle measurements confirmed that FPS is more hydrophobic than PS at the macroscopic level due to the introduction of fluorine. However, single molecule force spectroscopy experiments showed that the forces required to unfold the PS and FPS nanoparticles in water are indistinguishable, indicating that the strength of the hydrophobic effect that drives the self-assembly of PS and FPS nanoparticles is the same at the microscopic level. The divergence of hydrophobic effect at the macroscopic and microscopic level may hint different underlying mechanisms: the hydrophobicity is dominated by the solvent hydration at the microscopic level and the surface-associated interaction at the macroscopic level.

Hydrophobic interactions play an important role in many biochemical processes, including protein folding, membrane and micelle formation, and molecular recognition ([1–12]). Lum, Chandler, and Weeks first proposed the length-scale dependence theory of hydrophobic free energy (the LCW theory) ([13]). The hydration free energy scales with volume of hydrophobic solute at small regime while scales with surface area at lager regime ([14]). The former is an entropy-dominated process and the latter is an enthalpy-dominated process. After the LCW theory, researches shown that higher temperature decreased the hydration free energy while higher pressure increased the hydration free energy and reduced the crossover length ([15, 16]).

The development of single molecule force spectroscopy (SMFS) technique ([17–29]) promoted an experimental method to study hydrophobic interactions within hydrophobic polymers, such as polystyrene (PS) ([30]). Li and Walker investigated many hydrophobic dynamic processes by unfolding PS nanospheres by atomic force spectroscopy (AFM) ([31–33]). Additives such as NaCl and ethanol in water were also studied. NaCl lifted the plateau force of unfolding PS nanosphere and increased the hydration free energy while ethanol shown the opposite effects ([30]). Different substituted groups on PS shown that hydration free energy was affected by the size of monomer molecule ([31]). Later, Di et al. gave a direct experimental verification of the LCW theory based on the force-extension curve of unfolding PS nanospheres ([34]). Recently, Zeng and coworkers directly measured the hydrophobic force between two PS surfaces ([35]), and between AFM colloidal/drop probe and PS surface in NaCl solution ([36]) where the electrostatic force and bridging nanobubbles were ruled out ([37–39]).

In recent decades, hydrophobic/oleophobic materials have drawn great research interest due to their expanding applications ([40–47]). According to Young's equation ([48]), the lower surface tension of the solid leads to a larger contact angle, which indicates a more hydrophobic/oleophobic surface. In as early as the 1950s, Zisman proposed that the surface free energy of the chemical group followed the order $-CF_3$ < $-CF_2H$ < $-CF_2$ < $-CH_3$ < $-CH_2$ ([49]). Nakamae and coworkers also reported that the -$CF_3$ group showed the lowest surface free energy among various chemical groups ([50]). Therefore, to be hydrophobic/oleophobic, various fluorinated groups were introduced into materials due to their relatively lower surface free energy. Considering that substituted groups can affect the hydrophobic interaction, we wondered the direct fluorine substitution of hydrogen on the benzene side of PS exerts what influence on the hydrophobicity of PS at the macroscopic level, especially at the microscopic level.

In this work, the SMFS method is used to study the influence on hydrophobicity brought by F-substitution at the molecular level. PS and fluorine-substituted polystyrene (FPS) are typical hydrophobic polymers. Driven by hydrophobic interactions, these polymers tend to collapse into compact nanospheres in water to reach an energetically stable state ([14]). By unfolding such nanospheres using single-molecule AFM, we can quantitatively measure the hydrophobic effect of PS and FPS at the microscopic level.

To unfold the nanoparticles of hydrophobic polymers, it is vital to stably attach the polymers on the substrate and the cantilever tip. The common method used to coat a polymer to the substrate is first dissolving the polymer in a solvent and then spin-coating it ([30]). However, this physical adsorption method often results in many disadvantages, including inhomogeneous modification, difficulty in controlling the concentration and instability after a longer time, and especially non- specific interactions in the system, which always cause uncontrollable difficulties in carrying out experiments.

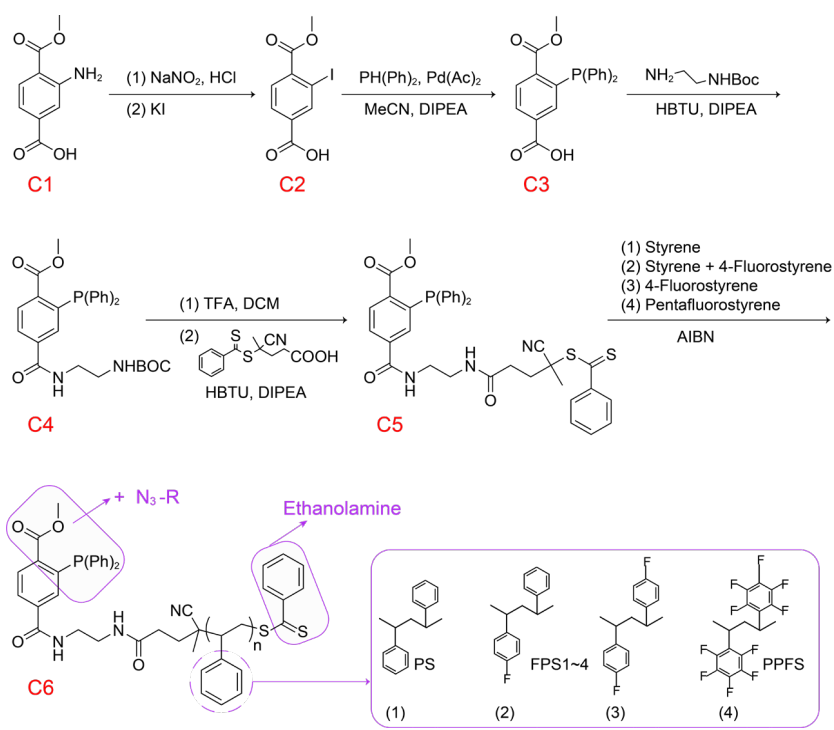

FIG. 1. The synthesis route of PS and several kinds of FPS polymers with bifunctional groups. The synthesis procedure was taken from *Ref.* ([51]) and our previous work ([34]) with some modifications. To be simple, each compound is named C with a number. C5 is the RAFT chain transfer agent. Different monomers from reaction C5 to C6 produce different target polymers. One terminal of polymer is a methyl ester group placed on the diphenylphosphine, and another terminal of polymer is a thiol group after the reduction reaction by ethanolamine

Therefore, to link the target polymer to the substrate via chemical linkage instead of unstable physical adsorption, we used reversible addition-fragmentation chain transfer (RAFT) polymerization method ([52, 53]) to synthesize the target polymer that possesses methyl ester placed on the diphenylphosphine and thiol group at two termini to establish chemical connection between the substrate and the AFM tip, respectively. Note that chemical linkage does not affect the following experiments because the handle is small enough compared with polymer itself and the handle is also hydrophobic. The synthesis route is shown in FIG. 1 ([34]) and the detailed procedure is shown in Supplementary Materials. We first synthesized the RAFT chain transfer agent, C5, through reactions C1 to C5. Then, we used the RAFT polymerization method to synthesize the target polymer, PS and FPS. Different monomer produces different polymer. We synthesized six polymers to investigate the difference in hydrophobicity brought by F-substitution at the microscopic level. They are PS, FPS1 (copolymer of 10% mol 4-fluorostyrene and 90% mol styrene at preset), FPS2 (copolymer of 40% mol 4-fluorostyrene and 60% mol styrene at preset), FPS3 (copolymer of 70% mol 4-fluorostyrene and 30%mol styrene at preset), FPS4 (poly-4-fluorostyrene), and PPFS (poly-pentafluorostyrene), respectively.

We carried out element analysis to determine the actual ratio of monomer of 4-fluorostyrene in the copolymer. The detailed information is listed in the Table I. We found that the actual ratio is smaller than the preset ratio, which indicates that fluorinated styrene is more difficult to be polymerized than styrene. Weight average molecular weight ($M_w$), number average molecular weight ($M_n$) and polydispersity were quantified by gel permeation chromatography (GPC). The detailed data is listed in Table II. At the same preset degree of polymerization, the actual molecular weight monotonically decreases from top to bottom in Table II, which also indicates that F-substitution makes polymerization difficult.

TABLE I. The ratio of 4-fluorostyrene in copolymer.

| Polymer | Presetting ratio | Actual ratio |
| --- | --- | --- |
| FPS1 | 10% | 6.25% |
| FPS2 | 40% | 25.2% |
| FPS3 | 70% | 40.5% |

TABLE II. The GPC results of the polymers.

| Polymer | $M_w$ | $M_n$ | polydispersity |
|---|---|---|---|
| PS | 86 388 | 61 723 | 1.40 |
| FPS1 | 74 433 | 51 795 | 1.44 |
| FPS2 | 68 069 | 36 342 | 1.87 |
| FPS3 | 54 945 | 32 932 | 1.67 |
| FPS4 | 51 487 | 30 758 | 1.67 |
| PPFS | 46 107 | 38 920 | 1.18 |

After the preparation of the polymer, we started to carry out SMFS experiments at room temperature (298 K) and 1 atm. The experimental scheme is illustrated in FIG. 2(a) Here, we take PS nanospheres as an example, and the other FPS polymers are similar. The methyl ester placed on the diphenylphosphine at one end of the polymer can react with the azide group modified on the substrate via Staudinger ligation ([54, 55]). Thus, the target polymer can be chemically linked to the substrate (For simplicity, it is not shown in FIG. 2(a) and see FIG. 1 for details). The other end of the polymer carried a thiol group after being reduced by ethanolamine. Driven by hydrophobic interactions, the hydrophobic polymer collapsed into a compact spheric structure. PS and PPFS attached to the monocrystalline silicon wafer in water did show nanosphere structure with a height of 2 ~ 8 nm by AFM imaging (FIG. 2(b) and (c)).

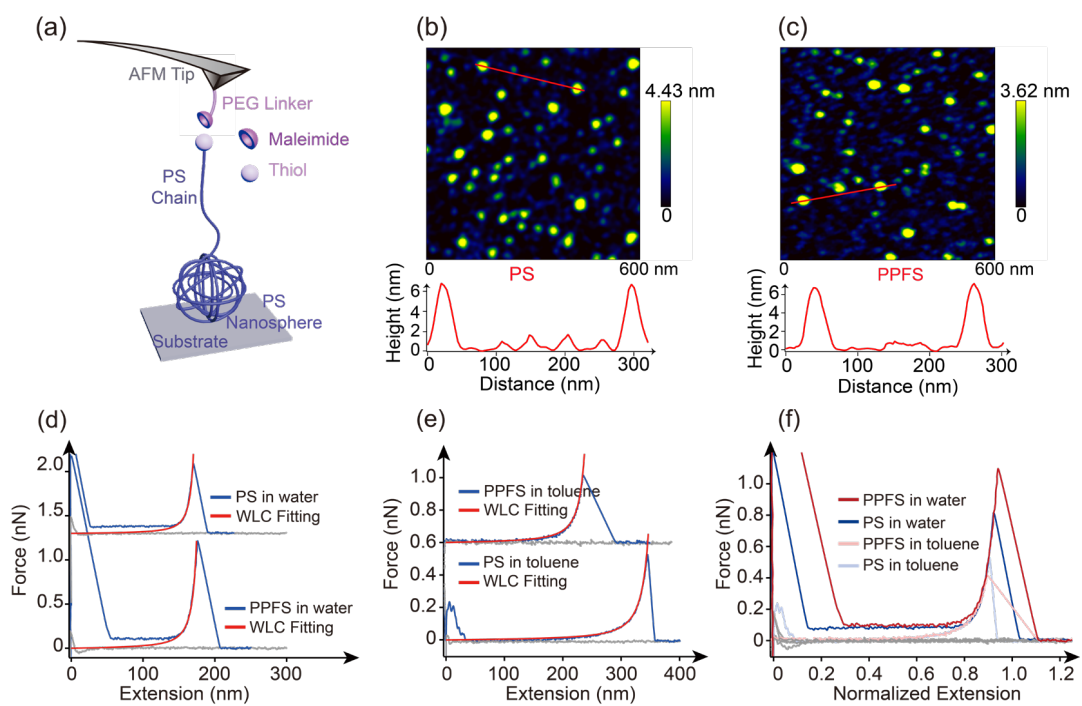

FIG. 2. The study of unfolding PS and FPS nanosphere by SMFS method. (a) Schematic illustration of unfolding PS or FPS nanospheres by AFM. Taking PS nanospheres as an example, the other FPS nanospheres are similar. (b) AFM image shows the PS polymer nanosphere structure with a height of 2 ~ 8 nm. (c) AFM image shows the PPFS polymer nanosphere structure with a height of 2 ~ 8 nm. (d) The force-extension curves of unfolding PS and PPFS nanospheres in water. (The curves of other FPS nanospheres are similar and shown in FIG. S7, S8, S9, and S10). The red curve is the WLC fitting to the elastic stretching part. (e) The force-extension curves of unfolding PS and PPFS nanospheres in the organic solvent toluene. The red curve is WLC fitting to the curve. (f) Normalized force-extension curves of unfolding PS and PPFS nanospheres by their contour length in water and toluene. The final elastic stretching regions overlap in all traces.

When the AFM tip approached the substrate, the thiol group had the chance to react with the maleimide group modified on the AFM tip. FIG. 2(d) shows the typical force-extension curves of unfolding PS and PPFS nanospheres in water (the force-extension curves of the other FPS polymers are similar and shown in Supplementary Materials, FIG. S7, S8, S9, and S10). They are similar to each other and feature a long force plateau with a subsequent rupture peak. The long force plateau indicates that the unfolding nanosphere is under a constant force before the collapsed nanosphere transforms into a fully extended chain. The rupture peak shows that the extended chain is further stretched until a certain chemical bond breaks. This elastic stretching process can be well fitted by the worm-like chain (WLC) model ([56, 57]). We also carried out the similar unfolding experiment in the organic solvent toluene instead of water. In toluene, PS and PPFS exist as random coils instead of collapsed nanospheres due to the lack of hydrophobic driving forces. The force-extension curve (FIG. 2(e)) shows a long elastic stretching process from the beginning that can also be well fitted by the WLC model. However, unlike FIG. 2(d), the abovementioned force plateau vanishes.

FIG. 2(f) shows the normalized force-extension curves in FIG. 2(d) and (e) by their contour length. The last elastic stretching parts overlap in all curves, indicating that not only the solvent but also the F-substitution does not affect the dynamic process of elastic stretching of the polymer chain. Furthermore, FIG. S11 shows the normalized force-extension curves of unfolding nanospheres of PS and all FPS polymers by their contour length. All the elastic stretching parts also overlap.

As it is commonly thought that the plateau force originates from overcoming the hydrophobic driving force, we use different FPS polymers to carry out unfolding experiments in water by AFM and measure the size of the plateau force. The detailed statistical data is listed in Table III. There is no difference among the plateau forces of PS and the five kinds of FPS polymers within errors. The amplitude of the plateau forces should be an intrinsic property. Based on the data mentioned above, we can conclude that the F-substitution on the phenyl group of PS does not influence the hydrophobicity of PS at the microscopic level. On the one hand, the force- extension curves of unfolding PS and FPS nanospheres show high consistency, and we cannot distinguish them. On the other hand, the hydrophobic free energy can be calculated just by the force-extension curve ([34]). Thus, there is no difference in the hydrophobic free energy between PS and FPS.

TABLE III. Unfolding plateau force of pulling different polymers in water.

| Polymer | Plateau Force (pN) (ave. ± std.) |
|---|---|
| PS | $106.0 \pm 2.8$ |
| FPS1 | $106.2 \pm 1.7$ |
| FPS2 | $109.2 \pm 3.6$ |
| FPS3 | 109.9 ± 2.8 |
| FPS4 | 106.3 ± 3.3 |
| PPFS | 106.9 ± 2.6 |

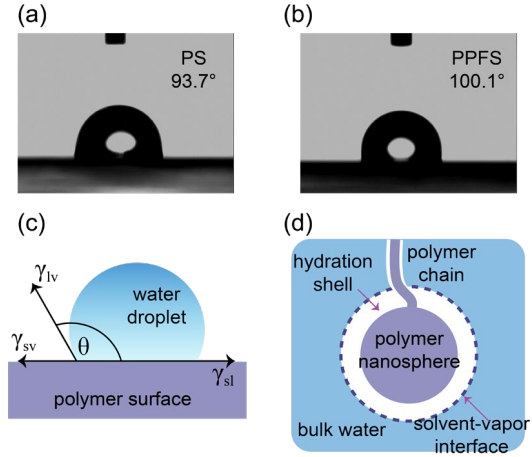

FIG. 3. Contact angle and hydration shell. (a) Water droplet on the PS surface. The contact angle is 93.7 ± 0.9°. (b) Water droplet on the PPFS surface. The contact angle is 100.1 ± 2.0°. The illustration for contact angle Young's equation. Here, $\gamma_{sl}$ represents interfacial tension or free energy between the liquid and the solid, $\gamma_{lv}$ the surface tension between the liquid and the vapor, $\gamma_{sv}$ the surface tension between the liquid and the vapor, and $\theta$ contact angle. (d) The illustration for hydration shell theory. There is an interface between polymer nanosphere and bulk water called hydration shell.

Next, we investigated the hydrophobic interaction strength of PS and FPS at the macroscopic level. We selected PPFS as the representative of FPS polymers because it has the highest content of fluorine. We measured the contact angle to quantify the difference in hydrophobicity between PS and PPFS. The contact angles of the water droplets on the PS and PPFS surfaces are 93.7 ± 0.9° and 100.1 ± 2.0° (FIG. 3(a) and (b)), respectively. There is a nearly 7% increase of the contact angle upon fluorination, indicating that PPFS is more hydrophobic than PS at the macroscopic level.

According to Young equation ([48]),

$$\gamma_{sv} - \gamma_{sl} = \gamma_{lv} \cos\theta , \qquad (1)$$

the contact angle is determined by three terms: solid surface tension ($\gamma_{sv}$), liquid surface tension ($\gamma_{lv}$), and interfacial free energy or surface tension ($\gamma_{sl}$) between the liquid and the solid (FIG. 3(c)). Therefore, the interaction between the liquid-solid surface is important to the contact angle at the macroscopic level. Based on Eq. 1, the interfacial free energy can be calculated and the detailed parameters of PS and PPFS are listed in Table IV. On account of contact surface area, the difference of energy of PS and PPFS is up to 24.2%.

TABLE IV. The surface parameters of PS and PPFS.

| Polymer | $\theta$ (deg) | $\gamma_{sv}$ (mJ/m$^2$) | $\gamma_{lv}$ (mJ/m$^2$) | $\gamma_{sl}$ (mJ/m$^2$) |
|---|---|---|---|---|
| PS | 93.7 | 40.6 [30] | 72.6 [30] | 45.3 |
| PPFS | 100.1 | 22.6 [58] | 72.6 | 35.3 |

However, at the microscopic level, there is no direct contact between the solid (PPFS or PS surface) and the liquid (water) ([3, 59]). According to the theory of hydration shell, there is a water-vapor interface between the polymer nanosphere and the bulk water ([60–62]). The unfolding of the nanosphere involves pulling the polymer chain through the water-vapor interface and then into bulk water. The entry of polymer chain into bulk water equals adding a certain surface area to the inner space of water, which makes the pulling forces proportional to the surface free energy of water. Therefore, the plateau force is determined by the water. The contrasting behaviors of hydrophobicity at the microscopic and the macroscopic level are due to their distinct underlying mechanism. At the microscopic level, it is the solvent-dominated effect, and at the macroscopic level, it is the surface-dominated effect. Previous works shown that substituted groups of benzene and tertiary butyl on PS affected the hydration free energy ([31]) which was different from F-substitution. The reason may lie in the bigger size of monomer changed the hydration process while fluorine was too small to influence the hydration process. This enlightened us that potential application of F-modified hydrophobic probe to study hydrophobic issues.

In summary, we used the SMFS method and the contact angle measurement to study the effects of F-substitution on the hydrophobicity of polymers at the microscopic and macroscopic levels, respectively. Intuitively, F-substitution would affect the hydrophobicity of PS at both the microscopic and macroscopic levels. However, quantitative force-extension curves showed the opposite. The indistinguishable plateau forces of PS and FPS polymers indicated that they have the same hydrophobic free energy at the microscopic level, although the contact angle measurement showed that

PPFS is more hydrophobic than PS at the macroscopic level. We inferred that at the microscopic level, due to the surface drying effect, the hydrophobic effect is mainly determined by the chemical properties of the solvent molecules. However, the surface dewetting effect is not applicable at the macroscale. At the macroscopic level, the chemical properties of the polymer play an important role and dominate the hydrophobic effects. Our study reveals the physical origin of hydrophobic effects at both the microscopic and the macroscopic levels, which may be important for the understanding of many phenomena, including the super hydrophobicity of lotus leaf, the ion transport in biological ion channels, and the performance of fluorinated surfaces at different length scales.

This research is supported mainly by the National Key R&D Program of China (Grant No.2020YFA0908100). We thank Dr. Zuoxiu Tie for GPC measurements and Dr. Ruixue Bai for NMR spectrum.

# Supplemental Material

# for

# Fluorination Increases Hydrophobicity at the Macroscopic Level but not at the Microscopic Level

Weishuai Di (邸维帅)[1], Xin Wang (王鑫)[1], Yanyan Zhou (周艳艳)[1], Yuehai Mei (梅岳海)[2], Wei Wang (王炜)[2,3,4*], and Yi Cao (曹毅)[1,2,3,4 §].

[1]*Oujiang Laboratory (Zhejiang Lab for Regenerative Medicine, Vision and Brain Health), Wenzhou Institute, University of Chinese Academy of Sciences, Wenzhou 325001, China*

[2]*Collaborative Innovation Center of Advanced Microstructures, National Laboratory of Solid State Microstructure, Department of Physics, Nanjing University, Nanjing 210093, China*

[3]*Institute for Brain Sciences, Nanjing University, Nanjing, 210093, China*

[4]*Chemistry and Biomedicine Innovation Center, Nanjing University, Nanjing 210093, China*

\* email:wangwei@nju.edu.cn

§ email:caoyi@nju.edu.cn

# I Synthesis Procedure

**C1→C2**: NaNO$_2$ (1.08 g, 15.66 mmol) in water (7.2 mL) and methyl-2-aminotetrphate (3.00 g, 15.37 mmol, **C1**) were slowly added to a round flask bottle dispersed in ice-cold concentrated HCl (30 mL) with vigorous stirring for 5 min to exclude the evolved gas. Then, the mixture was allowed to stir for 0.5 h, followed by filtering through a sintered filter funnel. The filtrate was directly added to a stirring solution of KI (25.80 g, 155.10 mmol) for 1 h to substitute the diazonium salt. Then, 96 mL dichloromethane (DCM) was added. The organic phase was washed with 24 mL saturated Na$_2$SO$_3$ solution, filtered through a polypropylene membrane, recrystallized from methanol (MeOH) and dried under vacuum to yield methyl-2-iodoterephthalate (1.80 g, 5.88 mmol, Yield = 38%, **C2**).

**C2 → C3**: The obtained methyl-2-iodorephathalate (3.00 g, 9.80 mmol, **C2**), Pd(Ac)$_2$ (22.4 mg, 0.10 mmol), 24 mL anhydrous acetonitrile (ACN) and 3.0 mL N,N-Diisopropylethylamine (DIPEA) (2.10 g, 16.24 mmol) were loaded in a flame-dried Schlenk tube and degassed through three freeze-pump-thaw cycles. Finally, diphenylphospine (1824.6 mg, 9.80 mmol) was injected through a rubber septum by a syringe under argon gas. The reaction mixture was allowed to reflux overnight, cooled to room temperature, diluted with DCM, washed by 1 M HCl, recrystallized from cold MeOH to obtain the desired **C3** (2.40 g, 6.56 mmol, Yield = 67%).

**$^1$H NMR** (500 MHz, CDCl$_3$), δ (ppm) = 8.06--8.10 (m, 2H), 7.66--7.67 (d, 1H, $J$ = 3.5 Hz), 7.27--7.39 (m, 10H), 3.75 (s, 3H).
**$^{13}$C NMR** (125 MHz, CDCl$_3$), δ (ppm) = 170.54, 166.71, 141.62, 141.39, 139.03, 138.88, 136.96, 136.87, 135.59, 133.94, 133.78, 131.84, 130.60, 129.73, 129.06, 128.71, 128.65, 52.38.
**ESI-MS (-)**, 364.1, [M]$^-$, cald for C$_{21}$H$_7$O$_4$P: 364.0864.

**C3 → C4**: **C3** (1.00 g, 2.74 mmol), O-Benzotriazole-N,N,N',N'-tetramethyl-uronium-hexafluorophosphate (HBTU) (1.24 g, 3.28 mmol), DIPEA (1020 μL, 5.48 mmol) and N-Boc-ethylenediamine (0.52 g, 3.28 mmol) were dissolved well in anhydrous DCM/ACN and stirred for 4 h at room temperature. The obtained mixture was subsequently washed with 1 M HCl, saturated NaHCO$_3$ solution and brine, dried by anhydrous MgSO$_4$ overnight. The crude product was purified by a flash chromatography purification (DCM : MeOH = 98 : 2, v/v) to obtain **C4** (1.18 g, 2.32 mmol, Yield = 85%).

**$^1$H NMR** (500 MHz, CDCl$_3$), δ (ppm) = 8.04--8.07 (dd, 1H, $J$ = 8.0, 3.9 Hz), 7.73--7.77 (dd, 1H, $J$ = 8.2, 1.2 Hz), 7.27--7.41 (m, 11H), 6.91--6.97 (br, 1H), 4.82--4.92 (br, 1H), 3.74 (s, 3H), 3.40--3.45 (m, 2H), 3.26--3.31 (m, 2H), 1.40 (s, 9H).
**$^{13}$C NMR** (125 MHz, CDCl$_3$), δ (ppm) = 166.67, 157.34, 141.61, 141.38, 137.22, 137.14, 136.94, 136.76, 136.61, 133.99, 133.83, 133.05, 130.78, 128.97, 128.67, 128.61, 126.37, 80.10, 53.44, 52.23, 41.94, 39.92, 28.33.
**ESI-MS (+)**, 545.3, [M+K]$^+$, cald for C$_{28}$H$_{31}$N$_2$O$_5$P: 506.1971.

**C4 → C5**: The obtained **C4** was deprotected by DCM/Trifluoroacetic acid (5mL, v/v = 3:1) at room temperature for 3 h until completely conversion as monitored by thin layer chromatography. After removing solvent under vacuum, the residue was dissolved in saturated NaHCO$_3$ solution in MeOH, filtered, and concentrated to yield **C5** precursor. To an anhydrous DCM (20 mL) solution of 4-cyano-4-(thiobenzoylthio) pentanoic acid (0.50 g, 1.79 mmol), HBTU (0.82 g, 2.15 mmol) and DIPEA (664 μL, 3.57 mmol) were added and dissolved well by stepwise addition of anhydrous ACN. Then **C5** precursor (0.87 g, 2.15 mmol) was added under argon flow. The reaction was conducted in an ice-water bath for 4 h to minimize the aminolysis of the dithiobenzoate group in alkaline environment. Then, the organic phase was extracted with 1 M HCl, saturated NaHCO$_3$ solution, brine and dried by anhydrous MgSO$_4$. After removal of the solvent, the crude product was loaded on a silica column eluting with DCM/MeOH (v/v = 99:1) to give **C5** as a red solid (0.50 g, 0.75 mmol, Yield = 42%).

**$^1$H NMR** (500 MHz, CDCl$_3$), δ (ppm) = 8.05--8.08 (dd, 1H, *J* = 8.0, 3.6 Hz), 7.86--7.89 (dd, 2H, *J* = 8.5, 1.0 Hz), 7.71--7.74 (dd, 1H, *J* = 8.0, 1.8 Hz), 7.54--7.58 (m, 1H), 7.27--7.40 (m, 13H), 6.75 (br, 1H), 6.16 (br, 1H), 3.73 (s, 3H), 3.48 (m, 2H), 3.42 (m, 2H), 3.35--3.60 (m, 6H), 1.91 (s, 3H).
**$^{13}$C NMR** (125 MHz, CDCl$_3$), δ (ppm) = 222.39, 171.75, 167.13, 166.68, 144.43, 141.83, 141.51, 137.15, 137.07, 136.63, 134.00, 133.83, 133.10, 132.98, 130.87, 129.00, 128.69, 128.60, 126.66, 126.44, 118.66, 99.99, 53.43, 52.26, 45.95, 40.88, 40.14, 33.95, 31.76, 24.32.
**ESI-MS (+),** 706.3, [M+K]$^+$, cald for C$_{36}$H$_{34}$N$_3$O$_4$PS$_2$ : 667.1728.

**C5 → C6**: **C5** is a Staudinger installed reversible addition-fragmentation chain transfer (RAFT) reagent controlling polymerizing monomers to form a bifunctional polymer chain with a Staudinger handle, which can selectively react with an azide group on the substrate, and a ditiobenzoate group on the other end, which can be furtherly transformed to thiol group to react with the maleimide on the AFM tip. Briefly, azodiisobutyronitrile (AIBN) (0.4 mg, 2.5 μmol) and **C5** (8.3 mg, 12.5 μmol) and freshly distilled styrene (1.30 g, 12.5 mmol, or other monomers with the same moles) were placed in a flame-dried Schlenk tube. Oxygen was removed by three freeze-pump-thaw cycles. The argon protected Schlenk tube was immersed in a 80 °C oil batch for 72 h to conduct the polymerization with a moderate conversion rate. After cooling to room temperature , Tetrahydrofuran was added to the tube to dissolve polymer. The diluted solution was dropwise added to a stirring cold MeOH solution to precipitate polymer for 3 times. The precipitate was collected and dried by vacuum.

The NMR characterization of six polymers (PS, FPS1~4, and PPFS) is as following (Fig. S1~S6):

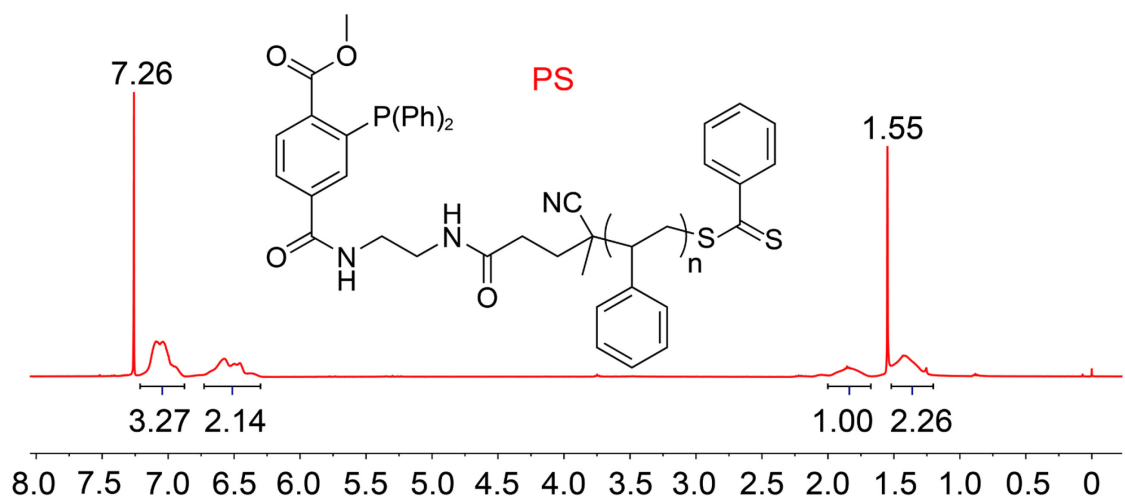

Fig. S1 The $^1$H NMR spectrum of PS.

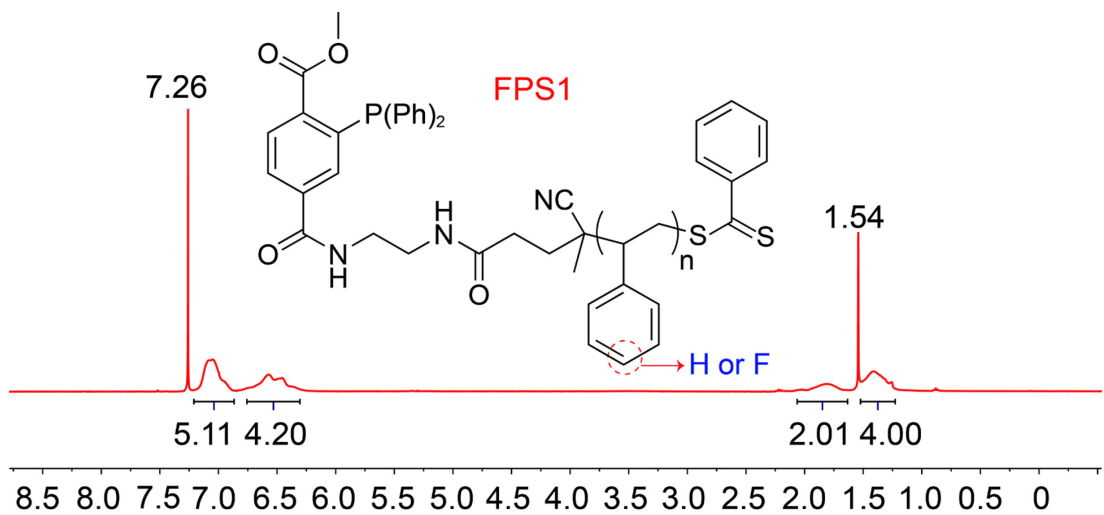

Fig. S2 The $^1$H NMR spectrum of FPS1.

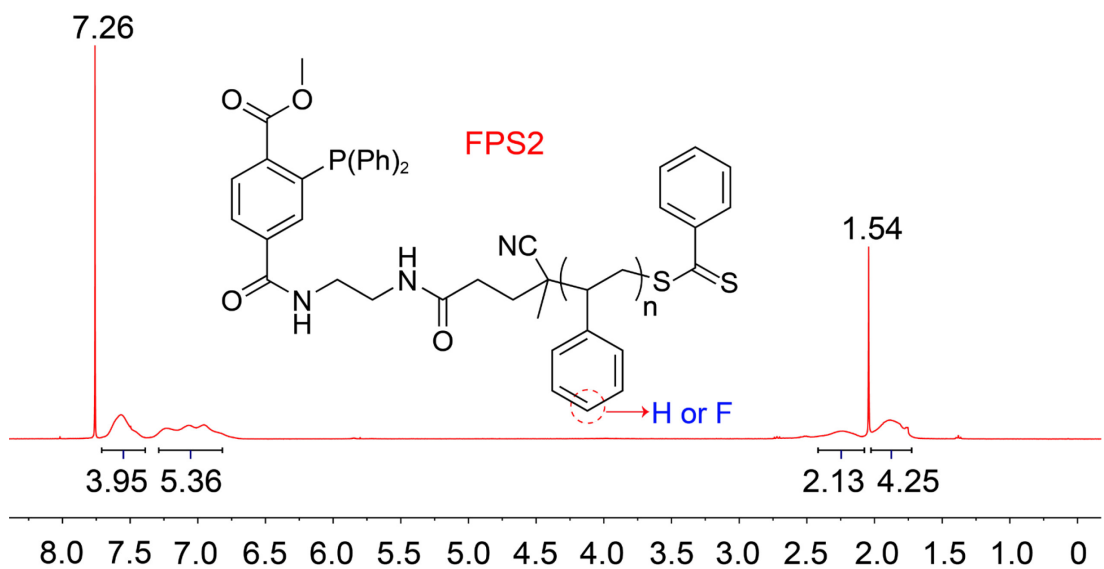

Fig. S3 The $^1$H NMR spectrum of FPS2.

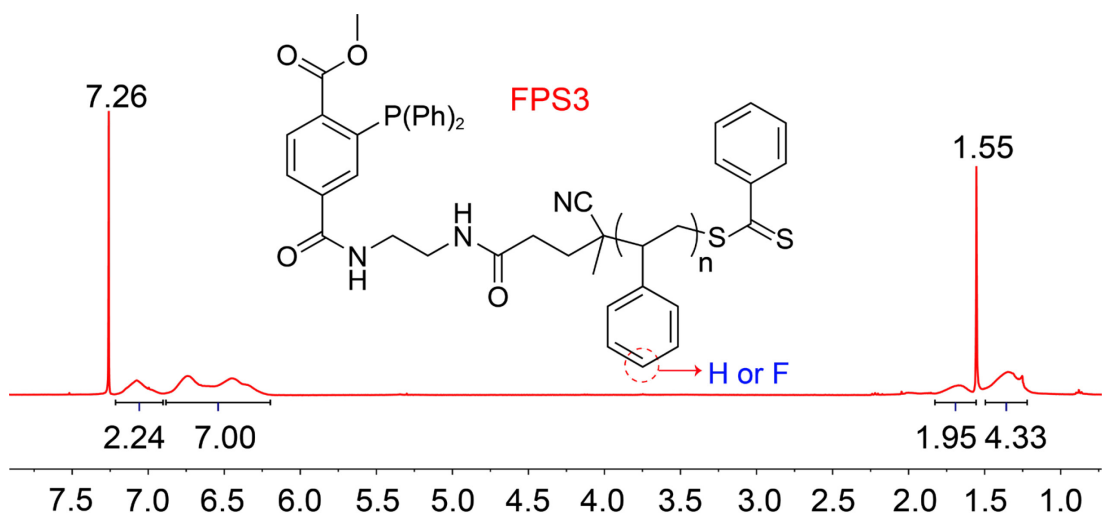

Fig. S4 The $^1$H NMR spectrum of FPS3.

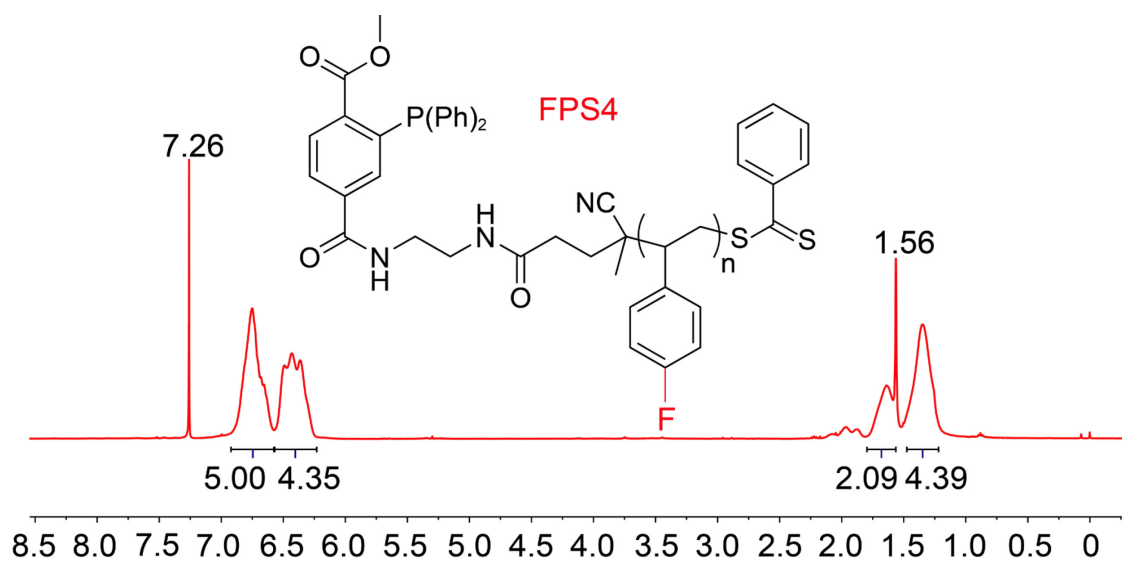

Fig. S5 The ¹H NMR spectrum of FPS4.

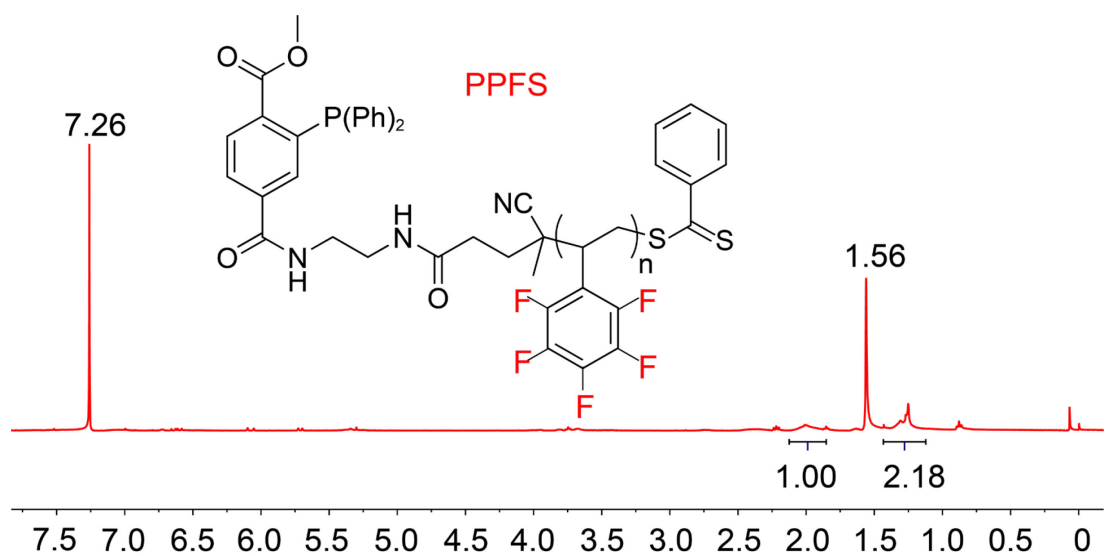

Fig. S6 The ¹H NMR spectrum of PPFS.

## II Force-extension Curves of Unfolding the Polymer Nnaosphere

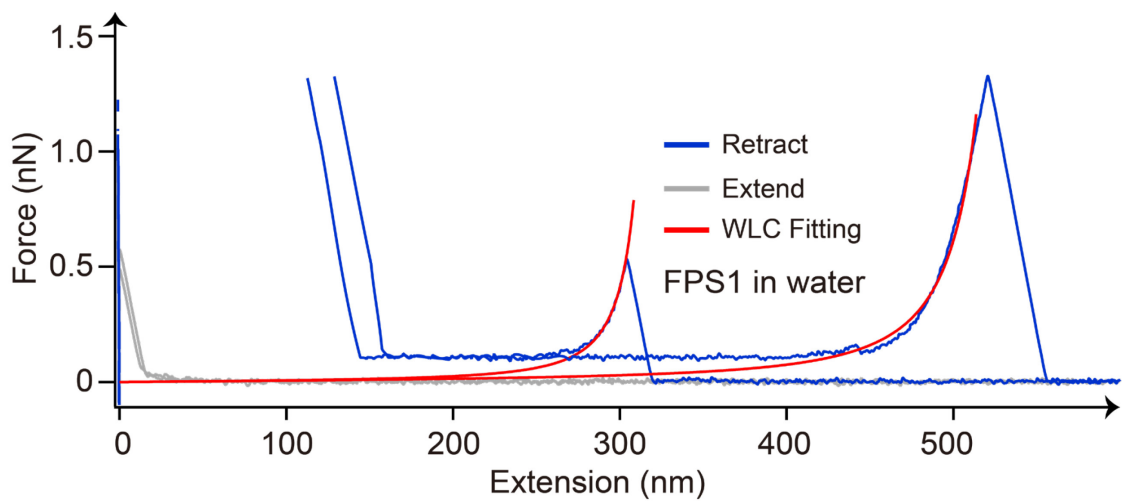

Fig. S7 The force-extension curves of unfolding FPS1 nanosphere in water. The grey line is extending curve, the blue line is retracting line, and the red line is WLC fitting to the elastic stretching of force-extension curves.

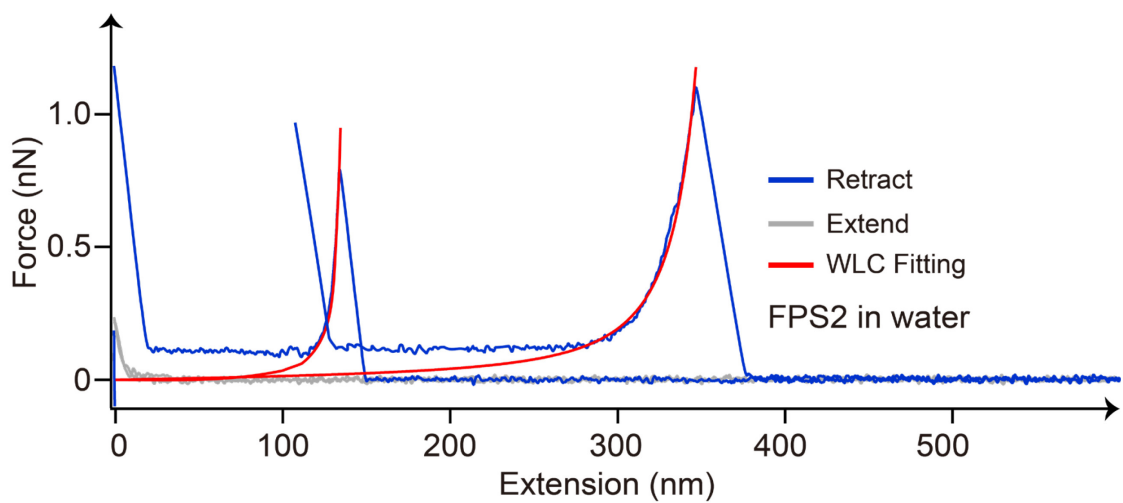

Fig. S8 The force-extension curves of unfolding FPS2 nanosphere in water. The grey line is extending curve, the blue line is retracting line, and the red line is WLC fitting to the elastic stretching of force-extension curves.

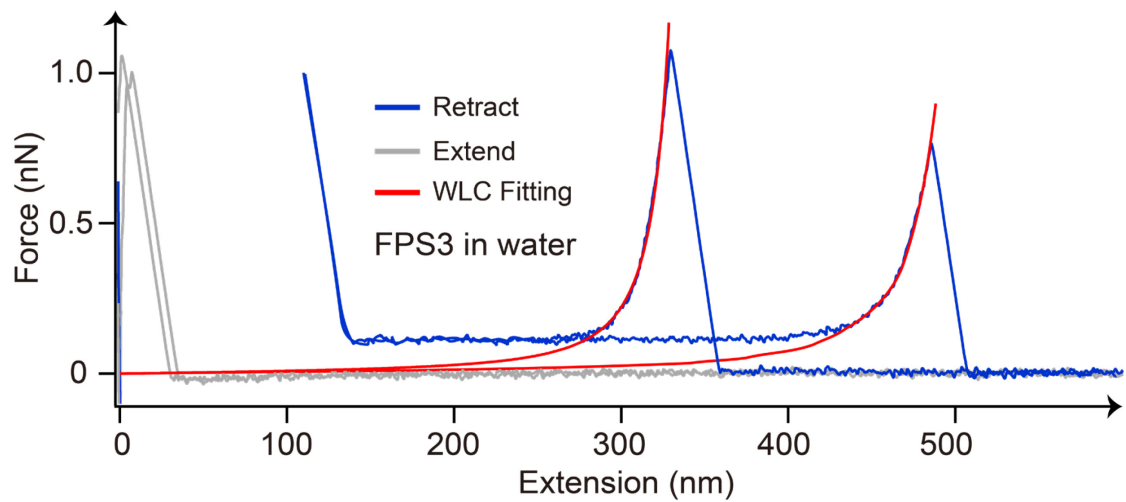

Fig. S9 The force-extension curves of unfolding FPS3 nanosphere in water. The grey line is extending curve, the blue line is retracting line, and the red line is WLC fitting to the elastic stretching of force-extension curves.

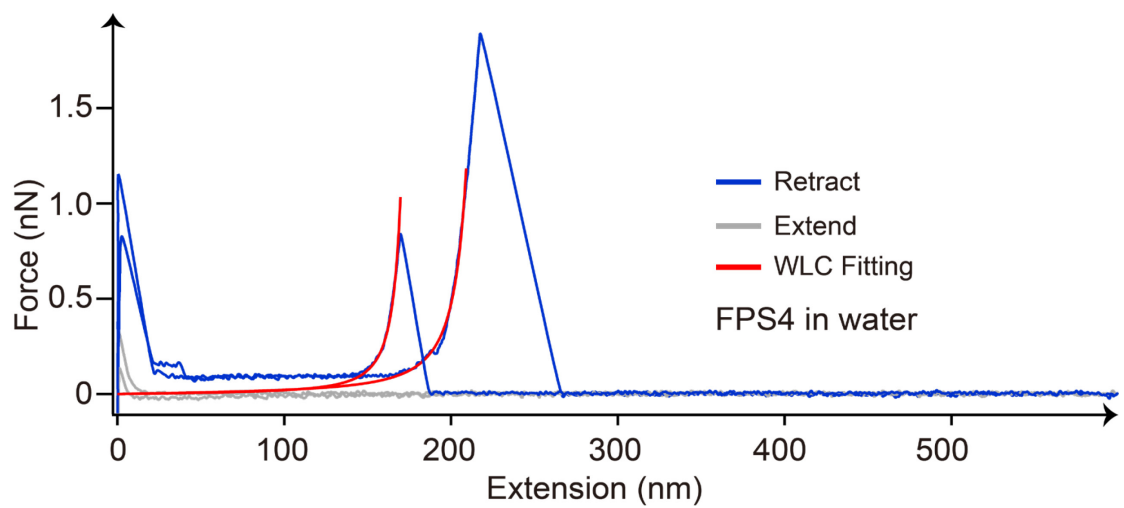

Fig. S10 The force-extension curves of unfolding FPS4 nanosphere in water. The grey line is extending curve, the blue line is retracting line, and the red line is WLC fitting to the elastic stretching of force-extension curves.

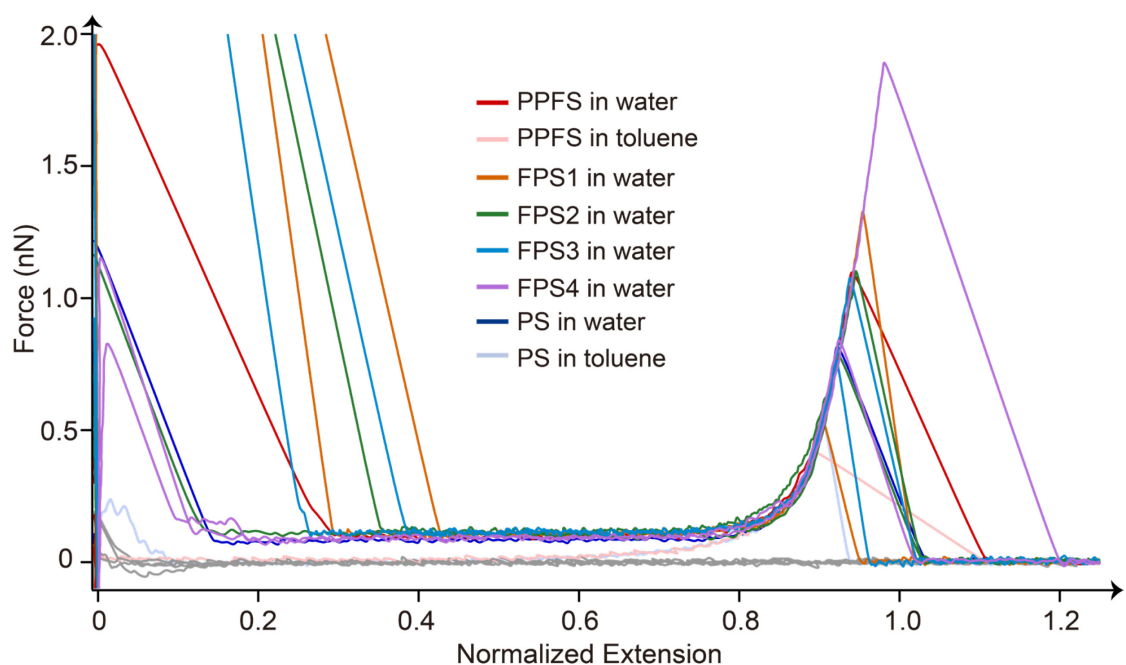

Fig. S11 The normalized force-extension curves of unfolding PS, FPS1, FPS2, FPS3, FPS4, and PPFS nanosphere in water and pulling PS and PPFS polymer chain in organic solvent toluene by their contour length. The elastic stretching parts of all curves are all overlapped. Note that these curves are from Fig. S7~S10 and Fig. 2 in main text.